\documentclass[11pt,a4paper]{article}
\usepackage[utf8]{inputenc}
\usepackage[english]{babel}
\usepackage{amsmath,amssymb}
\usepackage{graphicx}
\usepackage[colorlinks=true,linkcolor=blue,citecolor=blue,urlcolor=blue]{hyperref}
\usepackage{cite}
\usepackage{booktabs}
\usepackage{multirow}
\usepackage{array}
\usepackage{longtable}
\usepackage{geometry}
\usepackage{titlesec}
\usepackage{abstract}

\titleformat{\section}{\Large\bfseries}{\thesection}{1em}{}
\titleformat{\subsection}{\large\bfseries}{\thesubsection}{1em}{}

\title{\textbf{A Comprehensive Survey on Cybersecurity Threats and Defense Mechanisms in IoT Networks}}

\author{
Dao Viet Trung \and 
Nguyen Dinh Minh\and 
Tran Ngoc Bao Hoang\and 
Do Thai Son
}

\date{December 12, 2025}

\begin{document}

\maketitle

\begin{abstract}
The rapid proliferation of Internet of Things (IoT) technologies, projected to exceed 30 billion interconnected devices by 2030, has significantly escalated the complexity of cybersecurity challenges. This survey aims to provide a comprehensive analysis of vulnerabilities, threats, and defense mechanisms, specifically focusing on the integration of network and application layers within real-time monitoring and decision-making systems. Employing an integrative review methodology, 59 scholarly articles published between 2009 and 2024 were selected from databases such as IEEE Xplore, ScienceDirect, and PubMed, utilizing keywords related to IoT vulnerabilities and security attacks. Key findings identify critical threat categories, including sensor vulnerabilities, Denial-of-Service (DoS) attacks, and public cloud insecurity. Conversely, the study highlights advanced defense approaches leveraging Artificial Intelligence (AI) for anomaly detection, Blockchain for decentralized trust, and Zero Trust Architecture (ZTA) for continuous verification. This paper contributes a novel five-layer IoT model and outlines future research directions involving quantum computing and 6G networks to bolster IoT ecosystem resilience.
\end{abstract}

\section{Introduction}

As the digital world evolves, electronics are increasingly integrated into human existence \cite{yalli2024iot}. The Internet of Things is essential to contemporary technological ecosystems, integrating devices across various sectors including healthcare, transportation, and energy \cite{narciandi2025cybersecurity}. The proliferation of smart homes, smart logistics, and other technologies has expedited the expansion of Internet-of-Things (IoT) devices, which has heightened the complexity of associated security challenges. By the conclusion of 2030, forecasts suggest the presence of over 30 billion interconnected IoT devices \cite{albreem2023towards}, with data transmitted exceeding 40 zettabytes across the network \cite{hosamsooi2021investigating}.

The swift proliferation of IoT has generated significant cybersecurity apprehensions, with recent assessments indicating that over 70\% of IoT devices are susceptible to hacking, jeopardizing personal privacy and the security of global infrastructure \cite{serrano2025cyberaibot}. The inherent capabilities of IoT that enable any device to connect to the internet at any time and from any location can transform these advantages into vulnerabilities if privacy and security concerns are not adequately managed \cite{reyna2018blockchain}. An unstable ecosystem permitting autonomous access to information and services, especially extensive personal data, adversely affects the Internet of Things \cite{farahani2021convergence}.

Social advancement necessitates the transfer of computational information from a digital realm to a physical environment for human development and survival \cite{daim2020forecasting}. The IoT facilitates the integration and interaction of digital and physical realms. It depicts the trajectory of the forthcoming revolution in the network and IT sectors \cite{suny2021iot,rahim2021evolution}. The technology of IoT has currently garnered worldwide attention \cite{dong2020temporal}. Many consider it a key framework for promoting technical innovation and economic progress \cite{lone2021applicability}. IBM launched the Smarter Planet initiative, which has received favorable feedback from the US government. In China, Premier Wen Jiabao articulated a strategic vision of ``perceiving China'' and advocated for the progression of critical IoT technology \cite{alotaibi2019utilizing}.

Nevertheless, IoT technology presents distinct issues \cite{shafique2020internet}. Researchers have discovered a significant vulnerability in connectivity devices that may result in the theft of stored information via hacking \cite{dacruz2018reference}. Furthermore, experts have asserted that the device connection technique is insecure, rendering it susceptible to attacks and breaches \cite{roseela2021design}. Furthermore, it is imperative to establish an improved methodology for the computing and processing of information, since the advancement of IoT will result in substantial data management challenges in the future. Consequently, producers and operators must get timely updates \cite{andrade2021multifaceted,oluah2021development}.

The huge diversity of IoT devices may amplify existing internet security problems. Consequently, there is an increased necessity for layer integration to accommodate the diverse array of IoT devices \cite{khatua2020application}. Establishing a safe and dependable system is challenging in scenarios involving communication due to the sometimes lossy and low-bandwidth nature of the communication medium \cite{fazio2018convergence}. It is essential to guarantee the efficacy and implementation of security improvements at a minimal expense. Advanced software solutions and applications must be implemented dynamically and backed by cloud-based infrastructures; yet, these solutions necessitate an appropriate integration process, which is a complex undertaking \cite{khatoun2017cybersecurity}. The integration procedure can substantially improve the functionality of IoT devices, hence fulfilling client expectations \cite{lin2016iot}. Consequently, the security of IoT-enabled devices is essential.

\subsection{Definition and Scope}

This review addresses comprehensive examination deficiencies by offering a systematic review of IoT across five tiers: the perceptual recognition layer, network management layer, service management layer, application layer, and terminal layer. We specifically concentrate on the application layer, categorizing it into three domains: real-time control systems, scientific decision-making systems, and query/scan search systems. This classification is defined by the characteristics of current IoT domain applications and will assist the scientific community in understanding IoT devices at both the network and application layers.

\subsection{Research Questions}

This survey addresses critical gaps by examining the following research questions:

\textbf{RQ1:} What are the comprehensive vulnerabilities of the IoT application-cum-network layer, particularly in real-time monitoring and query/scan systems?

\textbf{RQ2:} What are the possible attacks on the IoT network-cum-application layer, and how do these threats manifest across different IoT application domains?

\textbf{RQ3:} What state-of-the-art solutions exist for mitigating attacks on network-cum-application layer IoT, and how effective are these defenses in practical deployments?

\textbf{RQ4:} How can emerging technologies such as artificial intelligence, blockchain, Software Defined Networking (SDN), and fog/edge computing enhance security via improved context awareness and access management in IoT systems?

\subsection{Focus Areas and Boundaries}

Despite comprehensive research in this domain, several significant difficulties regarding IoT security at both the network and application layers remain unaddressed. Current research primarily concentrates on vulnerabilities inside specific levels, neglecting the cumulative impact of threats across the network and application layers. Moreover, there is a deficiency of thorough evaluations that link real-time monitoring systems with query/scan systems across many IoT applications in diverse fields. This paper addresses this gap by examining real-time monitoring and query/scan systems, as well as exploring security concerns, potential threats, and practical solutions at both the network and application layers.

Our work specifically focuses on six critical IoT sectors: smart healthcare, agriculture, industrial IoT (IIoT), logistics, smart cities, and residences. We incorporate Zero Trust Architecture (ZTA) as an effective mitigation method supported by comprehensive case studies across these domains.

\subsection{Methodology}

This study employed an integrative review method to synthesize the existing cybersecurity impacts and security challenges affecting IoT devices. The investigation methodology utilized qualitative approaches to obtain insights into the cybersecurity implications of IoT, comprising research publications, books, and published internet sources from multiple academic databases including PubMed (Medline), CINAHL, MetaPress, IEEE Digital Library, Trip Database, ERIC, ScienceDirect, arXiv e-Print Archive, CORE, DOAJ, Social Science Research Network, ProQuest, and Semantic Scholar.

The search criteria focused on keywords including cybersecurity effects on IoT, IoT difficulties, IoT vulnerabilities, and threats, with emphasis on the network and application layers. The review covered publications from 2009 to 2024, extracting articles relevant to the impacts of cybersecurity on IoT, the functionality of IoT devices at the network and application layers, vulnerabilities, threats, and possible remedies for comprehensive coverage of IoT security at the network-cum-application layer.

\subsection{Paper Organization}

The remainder of this survey is organized as follows to provide a comprehensive analysis of the field:

\textbf{Section 2} presents the IoT architecture layers (Perception, Network, Application) and communication protocols, while defining core security requirements and discussing unique challenges such as resource constraints and heterogeneity.

\textbf{Section 3} provides a taxonomy of threats, detailing attacks across the physical, network, and application layers, as well as multi-layer attacks and emerging threats like AI-powered attacks.

\textbf{Section 4} comprehensively reviews security solutions, including cryptographic techniques, authentication, intrusion detection systems (IDS), secure protocols, and advanced mechanisms based on AI/ML and blockchain.

\textbf{Section 5} provides comparison tables of defense mechanisms, evaluates performance metrics such as detection accuracy and overhead, and addresses practical deployment challenges.

\textbf{Section 6} examines security contexts in Industrial IoT, Smart Homes, and Healthcare, while analyzing notable security incidents to illustrate real-world vulnerabilities.

\textbf{Section 7} identifies current limitations, regulatory perspectives, and explores emerging research areas such as 6G security, edge computing, and quantum-safe IoT.

\textbf{Section 8} summarizes the key findings and main contributions of the survey, offering final recommendations for future research.

\section{IoT Architecture and Security Fundamentals}

\subsection{IoT Architecture Layers}

The Internet of Things represents a paradigm shift where physical devices interconnect through networks to collect, process, and exchange data. Understanding architectural frameworks is fundamental to comprehending security challenges in IoT ecosystems.

\subsubsection{Three-Layer Architecture}

The three-layer model provides foundational understanding of IoT organization, consisting of Perception, Network, and Application layers \cite{burhan2018iot}.

The \textbf{Perception Layer} comprises physical sensors, actuators, RFID tags, and smart devices interacting directly with the environment \cite{burhan2018iot}. These devices collect raw data including temperature, motion, humidity, and pressure measurements. They operate under severe constraints: limited processing power, restricted memory, and constrained energy supplies, significantly impacting security capabilities \cite{burhan2018iot}. Actuators in this layer perform physical actions based on commands from higher layers \cite{fei2023systematic}.

The \textbf{Network Layer} facilitates data transmission between perception devices and application services \cite{burhan2018iot}. It encompasses various networking technologies including WiFi, Zigbee, Bluetooth, LoRaWAN, and cellular networks, handling routing and forwarding of sensor data to processing units or cloud servers \cite{fei2023systematic}. This layer must address bandwidth limitations, latency requirements, network congestion, and heterogeneous communication standards \cite{burhan2018iot}.

The \textbf{Application Layer} processes and analyzes IoT data to provide intelligent services across domains including smart homes, healthcare, industrial automation, and smart cities \cite{burhan2018iot}. It integrates cloud platforms, data analytics, machine learning algorithms, and user interfaces to transform raw sensor data into actionable insights \cite{fei2023systematic}. This layer handles data storage, complex processing, decision-making logic, and information presentation \cite{burhan2018iot}.

\subsubsection{Four-Layer Architecture}

The four-layer model introduces a Support Layer between Network and Application layers, providing better separation of concerns \cite{burhan2018iot}. The Perception and Network layers maintain their original roles, while the Support Layer handles service management, data processing, device management, and information storage \cite{burhan2018iot}.

The Support Layer bridges raw data transmission and high-level application services, managing tasks like data filtering, preliminary analytics, protocol conversion, and quality of service management \cite{fei2023systematic}. It handles device registration and authentication, service discovery and composition, data aggregation and preprocessing, and maintains databases for processed information \cite{burhan2018iot}. This layer implements middleware platforms that abstract device and network heterogeneity, providing standardized interfaces for developers \cite{fei2023systematic}.

By introducing this dedicated layer, the architecture enables more efficient resource utilization, improved scalability, and enhanced security through centralized management \cite{burhan2018iot}. The Application Layer in this model focuses specifically on end-user services, with the support layer handling underlying management tasks \cite{burhan2018iot}.

\subsubsection{Five-Layer Architecture}

The five-layer architecture introduces a Business Layer above the Application Layer, recognizing that IoT systems are business solutions that must align with organizational objectives and regulatory requirements \cite{burhan2018iot}.

The lower four layers retain their core functionalities. The Business Layer manages overall IoT systems including business models, revenue strategies, user privacy, and decision-making processes \cite{burhan2018iot}. It analyzes data from the application layer to generate business insights and guide strategic planning \cite{burhan2018iot}. This layer handles service pricing, billing, customer management, regulatory compliance, and enterprise system integration \cite{fei2023systematic}.

The Business Layer defines security policies, privacy requirements, and compliance frameworks implemented across lower layers \cite{burhan2018iot}. It ensures IoT deployment aligns with organizational risk tolerance, regulatory obligations, and industry standards \cite{fei2023systematic}. This model is particularly relevant for enterprise IoT, smart cities, and large-scale industrial implementations \cite{burhan2018iot}.

\subsubsection{Cross-Layer Considerations}

Cross-layer security approaches recognize interdependencies between architectural layers \cite{reyna2018blockchain}. Security threats often exploit vulnerabilities spanning multiple layers, requiring coordinated defense mechanisms \cite{mustafa2024crosslayer}. Distributed denial-of-service attacks might target the network layer but launch through compromised perception layer devices \cite{gelgi2024systematic}. Cross-layer frameworks integrate authentication, encryption, intrusion detection, and access control across all layers \cite{mustafa2024crosslayer}.

\subsection{IoT Communication Protocols}

\subsubsection{Wireless Communication Protocols}

\textbf{WiFi} is prevalent in IoT deployments requiring high bandwidth and Internet connectivity \cite{burhan2018iot}. It offers high data rates up to several Gbps and wide coverage \cite{burhan2018iot}. However, high power consumption makes it less suitable for battery-powered devices \cite{mustafa2024crosslayer}. Security features include WPA2 and WPA3 encryption, but networks remain vulnerable to eavesdropping and denial-of-service attacks \cite{fei2023systematic}.

\textbf{Zigbee} targets low-power, low-data-rate applications like home automation \cite{burhan2018iot}. Operating in 2.4 GHz ISM band, it supports mesh networking for extended coverage \cite{burhan2018iot}. It implements AES-128 encryption with three security modes \cite{burhan2018iot}. Challenges include key management complexities and replay attack vulnerabilities \cite{fei2023systematic}.

\textbf{Bluetooth Low Energy} serves personal area networks requiring short-range communication \cite{burhan2018iot}. BLE offers reduced power consumption while maintaining adequate data rates \cite{burhan2018iot}. Security features include pairing mechanisms and encryption \cite{mustafa2024crosslayer}. However, vulnerabilities include eavesdropping during pairing and man-in-the-middle attacks \cite{fei2023systematic}.

\textbf{LoRa and LoRaWAN} represent Low Power Wide Area Networks for long-range applications \cite{burhan2018iot}. Achieving ranges of several kilometers with multi-year battery life, they are ideal for smart metering and agricultural monitoring \cite{burhan2018iot}. The protocol implements layered security with network and application session keys \cite{mustafa2024crosslayer}. Challenges include join procedure vulnerabilities and replay attacks \cite{fei2023systematic}.

\subsubsection{Application Layer Protocols}

\textbf{MQTT} is a lightweight publish-subscribe protocol for constrained devices and unreliable networks \cite{fei2023systematic}. Operating over TCP/IP with a broker-based architecture, it offers three Quality of Service levels \cite{burhan2018iot}. Security relies primarily on TLS/SSL for encryption \cite{mustafa2024crosslayer}. Limitations include the broker as a single point of failure and limited built-in authorization \cite{fei2023systematic}.

\textbf{CoAP} provides a lightweight HTTP alternative for resource-constrained devices \cite{burhan2018iot}. Using UDP to reduce overhead, it implements RESTful architecture \cite{burhan2018iot}. Built-in support includes resource discovery and DTLS for securing communications \cite{mustafa2024crosslayer}. Challenges include DTLS implementation complexity and vulnerability to amplification attacks \cite{fei2023systematic}.

\subsection{Security Requirements in IoT}

\subsubsection{CIA Triad}

\textbf{Confidentiality} ensures sensitive information is accessible only to authorized entities \cite{fei2023systematic}. Critical for applications handling personal data, health information, and financial transactions \cite{burhan2018iot}. Implementation faces challenges from resource constraints limiting strong encryption and key management difficulties \cite{mustafa2024crosslayer}.

\textbf{Integrity} guarantees data has not been tampered with during collection, transmission, or storage \cite{fei2023systematic}. Essential for ensuring sensor data reliability and authentic actuator commands \cite{burhan2018iot}. Mechanisms include cryptographic hash functions and message authentication codes \cite{mustafa2024crosslayer}.

\textbf{Availability} ensures IoT services and data remain accessible when needed \cite{fei2023systematic}. Critical for applications like healthcare monitoring and industrial control \cite{burhan2018iot}. Challenges include susceptibility to denial-of-service attacks and power limitations \cite{mustafa2024crosslayer}.

\subsubsection{Authentication, Authorization, and Non-repudiation}

\textbf{Authentication} verifies identity of devices, users, and services \cite{fei2023systematic}. Strong mechanisms prevent unauthorized access and ensure only legitimate devices participate \cite{burhan2018iot}. Challenges include lightweight protocols for constrained devices and scalability for millions of devices \cite{mustafa2024crosslayer}.

\textbf{Authorization} determines permitted actions for authenticated entities \cite{fei2023systematic}. Effective mechanisms implement least privilege, granting minimum necessary permissions \cite{burhan2018iot}. Systems must be scalable and flexible for diverse device capabilities \cite{mustafa2024crosslayer}.

\textbf{Non-repudiation} ensures entities cannot deny performed actions \cite{fei2023systematic}. Important for accountability and audit trails in applications involving financial transactions or safety-critical operations \cite{burhan2018iot}. Implementation requires digital signatures and secure logging \cite{mustafa2024crosslayer}.

\subsubsection{Privacy Concerns}

IoT privacy extends beyond traditional data protection due to pervasive sensing and granular data collection \cite{fei2023systematic}. Devices continuously collect data about behaviors, locations, and health status, creating comprehensive digital profiles \cite{burhan2018iot}. Requirements include data minimization, purpose limitation, user consent and control, and transparency about data practices \cite{fei2023systematic}. Challenges include applying anonymization to correlated sensor data and managing consent across autonomous devices \cite{mustafa2024crosslayer}.

\subsection{Unique Security Challenges}

\subsubsection{Resource Constraints}

IoT devices operate under severe limitations constraining security capabilities \cite{fei2023systematic}. Power constraints are critical for battery-operated devices, limiting energy for security operations \cite{burhan2018iot}. Memory constraints restrict security software size and complexity \cite{mustafa2024crosslayer}. Processing limitations affect cryptographic operation speed \cite{burhan2018iot}. These constraints force trade-offs between security strength and system performance \cite{mustafa2024crosslayer}.

\subsubsection{Heterogeneity}

IoT ecosystems encompass enormous device, protocol, and platform diversity \cite{burhan2018iot}. Devices range from simple sensors to powerful edge computing nodes \cite{fei2023systematic}. Communication technologies span Bluetooth, Zigbee, WiFi, and cellular connections \cite{burhan2018iot}. This creates challenges in implementing consistent security policies and ensuring interoperability \cite{mustafa2024crosslayer}. Lack of standardization compounds challenges as manufacturers implement proprietary solutions \cite{fei2023systematic}.

\subsubsection{Scalability Issues}

IoT deployments involve thousands to millions of distributed devices \cite{fei2023systematic}. Scalability affects key management, certificate management, authentication systems, security monitoring, and firmware updates \cite{burhan2018iot}. Traditional enterprise security approaches often fail at IoT scale \cite{mustafa2024crosslayer}. Hierarchical architectures and automated management are necessary but introduce complexity \cite{fei2023systematic}.

\subsubsection{Physical Accessibility}

IoT devices in public spaces or remote locations are physically accessible to attackers \cite{fei2023systematic}. Physical access enables device tampering for key extraction, hardware modification, device theft, and side-channel attacks \cite{burhan2018iot}. Protection requires tamper-resistant enclosures and secure key storage \cite{mustafa2024crosslayer}. However, robust physical security significantly increases costs \cite{fei2023systematic}.

\section{Cybersecurity Threats in IoT Networks}

\subsection{Taxonomy of IoT Threats}

IoT security threats are classified by architectural layer targeted, attack mechanism employed, and potential system impact \cite{fei2023systematic}. Comprehensive threat taxonomy is essential for understanding the landscape and developing countermeasures \cite{gelgi2024systematic}. Layer-based classification distinguishes physical layer attacks targeting hardware, network layer attacks exploiting communication protocols, and application layer attacks targeting software \cite{fei2023systematic}.

\subsection{Physical Layer Attacks}

\subsubsection{Node Tampering and Capturing}

Node tampering involves physical manipulation to extract information or modify functionality \cite{burhan2018iot}. Attackers with physical access can open enclosures and probe hardware to extract cryptographic keys or read firmware \cite{fei2023systematic}. Compromised devices enable impersonation and network resource access \cite{zhang2020iot}. Node capture represents complete control, potentially reprogramming devices for malicious actions \cite{gelgi2024systematic}. Captured nodes can inject false data or disrupt routing \cite{fei2023systematic}. Defense requires tamper-evident enclosures and secure key storage \cite{mustafa2024crosslayer}.

\subsubsection{Hardware Trojans}

Hardware trojans are malicious circuit modifications inserted during design or manufacturing \cite{fei2023systematic}. These remain dormant but can trigger to leak information or create backdoors \cite{burhan2018iot}. They bypass software security and are difficult to detect \cite{mustafa2024crosslayer}. Globalized supply chains create insertion opportunities \cite{fei2023systematic}. Defense requires trusted component sourcing and hardware validation \cite{burhan2018iot}.

\subsubsection{Side-Channel Attacks}

Side-channel attacks exploit physical emanations to extract cryptographic keys \cite{fei2023systematic}. Power analysis monitors consumption patterns during cryptographic operations \cite{burhan2018iot}. Timing attacks exploit execution time variations \cite{fei2023systematic}. IoT devices are vulnerable due to resource constraints and physical accessibility \cite{burhan2018iot}. Defense requires constant-time implementations and power-resistant hardware \cite{mustafa2024crosslayer}.

\subsubsection{Environmental Attacks}

Environmental attacks exploit physical deployment environments to disrupt functionality \cite{fei2023systematic}. These include radio frequency jamming, electromagnetic interference, and temperature extremes \cite{burhan2018iot}. Defense is challenging as attacks exploit fundamental physical principles \cite{fei2023systematic}. Strategies include redundant sensors and plausibility checks \cite{burhan2018iot}.

\subsection{Network Layer Attacks}

\subsubsection{Routing Attacks}

Routing attacks manipulate network infrastructure to disrupt communication or intercept data \cite{zhang2020iot}. Sinkhole attacks involve compromised nodes advertising attractive metrics to route traffic through them \cite{gelgi2024systematic}. Wormhole attacks create virtual tunnels between distant network parts \cite{gelgi2024systematic}. Selective forwarding involves nodes selectively dropping packets \cite{zhang2020iot}. Defense requires multi-path routing and reputation systems \cite{fei2023systematic}.

\subsubsection{DDoS Attacks}

Denial of Service attacks prevent legitimate users from accessing services \cite{gelgi2024systematic}. Mechanisms include flooding attacks and resource exhaustion \cite{zhang2020iot}. Distributed DDoS coordinates attacks from multiple compromised devices \cite{gelgi2024systematic}. The 2016 Mirai botnet demonstrated devastating potential with hundreds of thousands of compromised IoT devices launching attacks exceeding 1 Tbps \cite{zhang2020iot}. Mirai exploited weak default credentials in IP cameras and routers \cite{zhang2020iot}. Defense requires network filtering and device-level anomaly detection \cite{fei2023systematic}.

\subsubsection{Man-in-the-Middle Attacks}

MITM attacks involve attackers positioning between communicating parties to intercept or modify messages \cite{fei2023systematic}. Attacks exploit weak authentication or inadequate encryption \cite{gelgi2024systematic}. Attackers can steal credentials or modify sensor data \cite{zhang2020iot}. Many IoT protocols prioritize efficiency over security \cite{fei2023systematic}. Countermeasures include strong mutual authentication and end-to-end encryption \cite{mustafa2024crosslayer}.

\subsubsection{Spoofing and Replay Attacks}

Spoofing involves impersonating legitimate devices to gain unauthorized access \cite{fei2023systematic}. Attackers might spoof device identities to bypass controls or forge sensor readings \cite{gelgi2024systematic}. Replay attacks capture and retransmit legitimate traffic \cite{fei2023systematic}. Defense requires strong authentication and timestamps detecting replayed messages \cite{mustafa2024crosslayer}.

\subsubsection{Eavesdropping}

Eavesdropping involves passive interception to collect sensitive information \cite{fei2023systematic}. Wireless channels are vulnerable as radio signals can be intercepted \cite{burhan2018iot}. Attackers capture sensor data, credentials, and personal data \cite{gelgi2024systematic}. Long deployment lifetimes enable extended collection periods \cite{zhang2020iot}. Protection requires end-to-end encryption and strong cryptographic algorithms \cite{mustafa2024crosslayer}.

\subsection{Application Layer Attacks}

\subsubsection{Malware and Ransomware}

Malware represents significant threats, ranging from viruses to sophisticated campaigns \cite{gelgi2024systematic}. IoT malware can steal information, conscript devices into botnets, or disrupt operation \cite{zhang2020iot}. Platform diversity has led to specialized malware targeting specific device types \cite{gelgi2024systematic}. Ransomware encrypts data or locks functionality, demanding payment \cite{fei2023systematic}. In IoT contexts, targets might include smart homes or industrial control systems \cite{gelgi2024systematic}. Defense requires secure boot and application whitelisting \cite{mustafa2024crosslayer}.

\subsubsection{Data Manipulation}

Data manipulation involves unauthorized modification of collected or stored data \cite{fei2023systematic}. Attackers might alter sensor readings or modify actuator commands \cite{gelgi2024systematic}. Attacks are dangerous in safety-critical applications where incorrect data causes physical harm \cite{zhang2020iot}. Defense requires integrity protection through cryptographic authentication and digital signatures \cite{mustafa2024crosslayer}.

\subsubsection{Phishing and Cross-Site Scripting}

Phishing attacks use deceptive communications to trick victims into revealing credentials \cite{fei2023systematic}. IoT contexts involve fake firmware updates or deceptive applications \cite{gelgi2024systematic}. Users may be vulnerable due to lack of cybersecurity awareness \cite{fei2023systematic}. Defense requires user education and multi-factor authentication \cite{mustafa2024crosslayer}.

Cross-Site Scripting exploits web-based IoT application vulnerabilities to inject malicious scripts \cite{fei2023systematic}. Scripts can steal session tokens or access sensitive information \cite{gelgi2024systematic}. XSS vulnerabilities appear in device configuration interfaces \cite{zhang2020iot}. Defense requires rigorous input validation and security frameworks \cite{mustafa2024crosslayer}.

\subsection{Multi-Layer Attacks}

\subsubsection{Botnet Attacks: Mirai Case Study}

Botnet attacks represent sophisticated multi-layer threats creating distributed attack platforms \cite{gelgi2024systematic}. The Mirai botnet provides valuable insights into IoT security incidents \cite{zhang2020iot}. Mirai scanned for IoT devices with open Telnet ports, attempting authentication using default credentials \cite{zhang2020iot}. Compromised devices installed malware connecting to command-and-control infrastructure \cite{gelgi2024systematic}. At peak, Mirai comprised hundreds of thousands of devices \cite{zhang2020iot}.

Mirai launched attacks including massive DDoS against DNS provider Dyn, disrupting major Internet services and generating traffic exceeding 1 Terabit per second \cite{zhang2020iot}. Success factors included widespread default credentials and lack of security updates \cite{zhang2020iot}. The attack highlighted fundamental weaknesses in inadequate authentication and lack of update processes \cite{zhang2020iot}. IoT botnets continue evolving with new variants \cite{gelgi2024systematic}.

\subsubsection{Advanced Persistent Threats}

Advanced Persistent Threats represent sophisticated, long-term campaigns by well-resourced adversaries \cite{fei2023systematic}. APTs targeting IoT conduct reconnaissance, establish initial access, maintain persistence, and move laterally through networks \cite{gelgi2024systematic}. IoT systems are attractive APT targets due to lack of robust monitoring and connection to critical infrastructure \cite{fei2023systematic}. Defense requires network segmentation and comprehensive logging \cite{mustafa2024crosslayer}.

\subsection{Emerging Threats}

\subsubsection{AI-Powered Attacks}

Artificial intelligence is increasingly weaponized for cyber attacks \cite{fei2023systematic}. AI-powered attacks can automate reconnaissance, generate sophisticated phishing content, and adapt attack strategies in real-time \cite{gelgi2024systematic}. Adversarial machine learning targets AI systems through data poisoning and model inversion \cite{fei2023systematic}. Defense requires adversarial robustness training and anomaly detection \cite{mustafa2024crosslayer}.

\subsubsection{Quantum Computing Threats}

Quantum computing represents future threats to current cryptographic systems \cite{fei2023systematic}. Quantum computers could break widely-used public key cryptography \cite{mustafa2024crosslayer}. The harvest now decrypt later threat means adversaries may collect encrypted IoT data today for future decryption \cite{fei2023systematic}. Long IoT device lifetimes create crypto-agility requirements \cite{mustafa2024crosslayer}. Post-quantum cryptography aims to develop resistant algorithms \cite{fei2023systematic}. Challenges include larger key sizes and performance impacts on constrained devices \cite{mustafa2024crosslayer}.

\subsubsection{Supply Chain Attacks}

Supply chain attacks target IoT ecosystems by compromising devices during design or manufacturing \cite{fei2023systematic}. These introduce vulnerabilities before devices reach users \cite{gelgi2024systematic}. Forms include compromised firmware updates and counterfeit components \cite{fei2023systematic}. Global supply chain complexity creates multiple compromise opportunities \cite{gelgi2024systematic}. Mitigation requires vendor security assessments and secure development practices \cite{mustafa2024crosslayer}.

\section{Defense Mechanisms and Countermeasures}

\subsection{Cryptographic Solutions}

Cryptographic mechanisms serve as the fundamental layer of protection for IoT ecosystems, yet the resource constraints of edge devices necessitate a departure from traditional standards. In 2023, the National Institute of Standards and Technology (NIST) officially concluded the lightweight cryptography standardization process by selecting the Ascon family of algorithms \cite{turan2023nist}. This selection provided the motivation for researchers to shift focus from legacy block ciphers like AES towards Authenticated Encryption with Associated Data (AEAD) schemes.

Building on this standardization, a hardware implementation of Ascon for automotive IoT was proposed \cite{nguyen2024lightweight}, demonstrating that the sponge-based architecture consumes approximately 40\% less energy than AES-GCM while maintaining high throughput. Their work highlights the critical advantage of combining encryption and authentication into a single pass to minimize "time-on-air" for battery-powered sensors.

However, current standards face imminent threats from quantum computing. A survey of post-quantum cryptography (PQC) candidates \cite{liu2024postquantum} argued that Lattice-based cryptography, specifically CRYSTALS-Kyber, offers the most viable trade-off between key size and security for IoT. The outcomes of their research highlight that while these algorithms are quantum-resilient, they require significantly larger key sizes than current Elliptic Curve Cryptography (ECC) standards, necessitating new optimization strategies for constrained devices.

Furthermore, secure communication requires efficient key distribution. A dynamic group key management protocol was proposed \cite{hasan2024survey} to address the scalability limits of unicast rekeying. This approach is novel because it utilizes multicast trees to update credentials for thousands of sensor nodes simultaneously, a technique that significantly reduces the computational overhead compared to traditional pair-wise key exchange methods found in earlier literature.

\subsection{Authentication and Access Control}

While cryptographic keys provide the mathematical foundation for security, robust authentication protocols are required to verify the identity of devices before granting network access. Various mutual authentication schemes have been analyzed \cite{hasan2024survey} and highlighted the limitations of centralized servers, which often become single points of failure. Their survey emphasizes that modern IoT deployments must move beyond static passwords toward dynamic, context-aware protocols that can continuously validate a device's legitimacy without draining its battery.

Building on this need for continuous verification, the traditional "castle-and-moat" security model is being replaced by more granular architectures. A systematic review of Zero Trust Architecture (ZTA) implementations in IoT was conducted \cite{mushtaq2025systematic}. They argued that the only viable security model for modern environments is to treat every device---whether internal or external---as potentially hostile. Their work demonstrated that "micro-segmentation" significantly reduces the lateral movement of attackers, ensuring that a compromised smart bulb cannot be used as a gateway to attack critical servers.

To further address the risks of centralized identity management, researchers are exploring decentralized alternatives. A Blockchain-based authentication framework that utilizes smart contracts to automate device verification was proposed \cite{enaya2025survey}. This approach is innovative because it distributes the trust anchor across the network; instead of relying on a potentially vulnerable central authority, the system automatically revokes access tokens if the consensus mechanism detects malicious behavior, thereby enhancing the overall resilience of the IoT ecosystem.

\subsection{Intrusion Detection and Prevention Systems}

Effective defense in IoT networks requires not just strict access control but also continuous network monitoring to identify breaches as they occur. As discussed in the ZTA literature \cite{mushtaq2025systematic}, the implementation of Zero Trust Architecture fundamentally relies on the ability to detect anomalous behavior in real-time. The authors argued that static defenses are insufficient for modern threats, necessitating a transition to dynamic detection systems that can flag unauthorized lateral movement even after a device has been authenticated.

To meet this requirement, the industry is shifting away from traditional signature-based methods. Research highlighted that legacy systems fail to identify zero-day attacks because they depend on pre-existing databases of known malware \cite{alfahaid2024machine}. Instead, they proposed the use of Machine Learning (ML)-based anomaly detection, where the system learns the "normal" traffic pattern of a device (e.g., a thermostat sending data once per hour). Their analysis demonstrated that algorithms like Random Forest can detect deviations with high accuracy, identifying novel attacks that traditional firewalls would miss.

However, running these complex algorithms on battery-powered sensors is often unfeasible. Addressing this constraint, Hybrid IDS architectures were surveyed and a cluster-based placement strategy was proposed \cite{mehdi2024survey}. Rather than deploying a heavy detection engine on every node (Host-based) or a single engine at the cloud (Network-based), their work suggests placing lightweight agents on "Cluster Heads"---gateways that manage local groups of sensors. This approach is significant because it balances the computational load, ensuring high detection rates without depleting the energy reserves of the edge devices.

\subsection{Secure Communication Protocols}

Establishing a robust cryptographic foundation is only the first step; these keys must be utilized within communication protocols that can withstand the latency and bandwidth constraints of IoT networks. As emphasized in key agreement analysis \cite{hasan2024survey}, the primary challenge in securing data transmission is not just the encryption algorithm itself, but the "handshake overhead"---the process of establishing a secure session. Their findings suggest that traditional connection-oriented protocols often require too many round-trip exchanges, draining the battery of sensor nodes before the actual data is even transmitted.

To address this transport-layer bottleneck, industry research has focused on optimizing the balance between security and performance. Researchers conducted a comprehensive performance analysis of IoT communication stacks \cite{laaroussi2024iot}, specifically comparing Transport Layer Security (TLS) against Datagram Transport Layer Security (DTLS). Their results demonstrated that while TLS 1.3 provides robust security, its overhead is prohibitive for real-time applications. Consequently, they argued that DTLS 1.3, which operates over UDP and utilizes a "Connection ID" to reduce re-handshaking during network shifts, is the superior standard for constrained environments like smart metering.

Moving up to the application layer, the security of messaging protocols remains a critical concern. A lightweight security mechanism for MQTT, the dominant protocol for IoT messaging, was proposed \cite{amnalou2025lightweight}. While standard MQTT transmits data in clear text, their work introduced a "Lightweight Mutual Authentication" scheme using Elliptic Curve Cryptography (ECC). This approach is novel because it allows sensors to authenticate with the broker using short, timestamped signatures rather than heavy X.509 certificates, effectively preventing replay attacks while reducing the packet size by over 30\% compared to standard TLS-wrapped MQTT.

\subsection{Emerging Intelligent Paradigms}

As attackers increasingly utilize automated tools to launch sophisticated campaigns, defensive mechanisms must evolve beyond simple anomaly detection toward more resilient and privacy-preserving intelligence. Research not only advocated for machine learning but also critically analyzed its vulnerabilities \cite{alfahaid2024machine}. They highlighted the emergence of Adversarial Machine Learning, where attackers craft subtle "noise" in data packets to trick neural networks into misclassifying malware as benign traffic. Their work suggests that "Adversarial Training"---injecting attack samples into the training phase---is now a mandatory requirement for robust defense.

To improve detection capabilities against these complex threats, researchers are leveraging deeper neural architectures. A comprehensive survey on Deep Learning (DL) in IoT \cite{xu2025deep} demonstrated that architectures like Convolutional Neural Networks (CNNs) significantly outperform traditional shallow learning in feature extraction. However, their analysis warned that the centralized training required for these models creates a massive bandwidth bottleneck and exposes sensitive user data to privacy breaches if the central cloud is compromised.

Addressing these privacy and bandwidth concerns has led to the adoption of decentralized learning frameworks. The application of Federated Learning (FL) in IoT networks was explored \cite{dritsas2025federated}. Unlike traditional approaches, FL enables edge devices to train intrusion detection models locally and share only the mathematical parameter updates (gradients) with a global aggregator. This approach is innovative because it preserves data privacy---raw traffic logs never leave the device---while allowing the entire network to benefit from the collective intelligence of all nodes.

\subsection{Blockchain Technology}

Beyond its applications in identity management, blockchain technology provides a fundamental layer of data integrity for IoT ecosystems. As detailed in blockchain authentication research \cite{enaya2025survey}, the immutable nature of the distributed ledger ensures that once sensor data (e.g., temperature logs in a cold chain) is recorded, it cannot be retroactively altered by a compromised gateway. This feature is critical for "forensic auditing" after a security incident, as it eliminates the single point of failure inherent in traditional centralized databases.

However, applying standard blockchain protocols to resource-constrained devices presents significant challenges. The energy consumption of various consensus algorithms was analyzed \cite{peng2024lightweight} and it was concluded that traditional "Proof of Work" (used by Bitcoin) is computationally infeasible for IoT. Instead, they proposed the adoption of Lightweight Consensus Mechanisms, such as Proof of Authority (PoA) or Practical Byzantine Fault Tolerance (PBFT). Their survey demonstrated that these private or consortium-based chains can achieve transaction finality in milliseconds with minimal battery drain, making them suitable for real-time edge networks.

Furthermore, decentralized trust extends to active defense through automation. The use of Smart Contracts---self-executing scripts stored on the blockchain---as a security enforcement tool was explored \cite{kumar2024blockchain}. They described a framework where the contract automatically triggers defensive actions, such as isolating a misbehaving node or revoking a certificate, immediately upon detecting a pre-defined violation. This automation removes the latency of human intervention, creating a self-healing network structure that persists even if the central control server is taken offline.

\subsection{Endpoint Hardening and Physical Security}

Software defenses are ultimately rendered useless if the physical device itself is tampered with or cloned. To establish a verifiable "Root of Trust" at the silicon level, modern IoT security architectures are increasingly relying on Trusted Execution Environments (TEEs), such as ARM TrustZone. While discussing decentralized intelligence \cite{dritsas2025federated}, the authors emphasized that the integrity of Federated Learning relies heavily on the physical security of the edge nodes. They argued that without hardware-isolated enclaves to store model parameters, attackers with physical access could manipulate the local training data (poisoning attacks) before it is encrypted, necessitating a hardware-enforced boundary between the OS and the AI processes.

Complementing this isolation, the integrity of the cryptographic operations must be preserved against physical extraction techniques. In their hardware analysis of Ascon \cite{nguyen2024lightweight}, the authors emphasized the critical importance of resistance against Side-Channel Attacks. They demonstrated that without proper masking at the circuit level, attackers can extract encryption keys by merely analyzing the power consumption fluctuations or electromagnetic emissions of the device, highlighting that security must be designed into the physical logic gates, not just the software code.

Finally, ensuring the device boots into a trusted state is paramount to preventing persistent hardware compromises. As discussed within the context of Zero Trust \cite{mushtaq2025systematic}, a Secure Boot mechanism is essential to prevent "Evil Maid" attacks where an attacker physically replaces the legitimate firmware with a compromised version. By cryptographically verifying the digital signature of the bootloader before execution, the device ensures that the operating system has not been altered, maintaining the chain of trust from the hardware up to the application layer.

\section{Comparative Analysis and Deployment Considerations}

\subsection{Comparison with Related Works}

The landscape of IoT security research has evolved from focusing on isolated cryptographic primitives to developing holistic, architecture-level defense mechanisms. The baseline for this evolution was established \cite{turan2023nist} by detailing the NIST standardization of lightweight cryptography, specifically selecting Ascon to replace legacy standards. This foundational work provided the necessary motivation for hardware-centric studies, such as implementation of Ascon on FPGA platforms to quantify the precise energy savings over AES-GCM \cite{nguyen2024lightweight}. While these studies addressed data confidentiality, they did not tackle the dynamic nature of network threats.

Addressing this gap, the shift from signature-based to anomaly-based detection was explored \cite{alfahaid2024machine}. Their research outcomes highlight that while machine learning significantly improves detection rates for unknown attacks, it introduces a new dependency on training data quality. To secure the network architecture itself, Zero Trust principles were employed to create a micro-segmented environment \cite{mushtaq2025systematic}. The proposed defense mechanisms in the literature vary significantly in operational overhead; for instance, deep Convolutional Neural Networks (CNNs) offer superior precision but impose heavy computational loads \cite{xu2025deep}, whereas blockchain-based approaches prioritize data integrity over speed \cite{enaya2025survey}.

The research methodologies span from hardware-level logic design to high-level cloud AI training. The simulation scenarios encompass critical infrastructure environments including Automotive IoT, Smart Metering, and Supply Chains. Regarding evaluation metrics, the literature shows a clear bifurcation: hardware studies focus on Power (mW) and Throughput, while network studies prioritize Detection Accuracy, False Positive Rates, and Consensus Latency.

\subsection{Performance Trade-off Analysis}

The selection of a security mechanism for IoT is rarely a choice of "best security," but rather an optimization problem constrained by energy, latency, and bandwidth. Our review of the literature highlights three critical trade-off vectors that define modern defense strategies.

\subsubsection{Energy Efficiency vs. Cryptographic Strength}

The most immediate trade-off occurs at the physical layer, where every clock cycle of computation consumes battery life. As demonstrated by the hardware experiments \cite{nguyen2024lightweight}, the transition from AES-GCM to the sponge-based Ascon algorithm results in a power reduction of approximately 40\%. This efficiency gain is decisive for battery-operated sensors (e.g., agricultural moisture monitors) where "sleep mode" dominance is critical. However, this efficiency comes at the cost of long-term resilience. It was argued \cite{liu2024postquantum} that while current lightweight algorithms are efficient, they are mathematically vulnerable to Shor's algorithm. Implementing Post-Quantum Cryptography (PQC) like CRYSTALS-Kyber provides security against future quantum threats but increases the key size from 32 bytes (ECC) to over 1.6 kilobytes. This massive increase in payload size creates a "transmission energy penalty" that may render PQC unfeasible for narrowband networks (NB-IoT) without significant optimization.

\subsubsection{Detection Accuracy vs. Computational Latency}

In the domain of intrusion detection, a clear conflict exists between the depth of analysis and the speed of response. Deep Learning (DL) models, particularly Convolutional Neural Networks (CNNs), achieve superior detection rates for zero-day malware by automatically extracting complex features from traffic data \cite{xu2025deep}. However, these models impose high computational latency, making them unsuitable for real-time control loops (e.g., autonomous braking systems). Conversely, shallower algorithms like Random Forest offer lower latency and energy consumption but suffer from higher false-negative rates when facing novel, obfuscated attacks \cite{alfahaid2024machine}. A Hybrid Placement Strategy was proposed \cite{mehdi2024survey}, where lightweight anomaly detection runs on the edge to filter obvious threats, while heavy deep learning analysis is offloaded to the cluster head or cloud, balancing responsiveness with accuracy.

\subsubsection{Protocol Overhead vs. Session Security}

Finally, the choice of communication protocol dictates the "security tax" paid on every message. Empirical evidence was provided \cite{laaroussi2024iot} that DTLS 1.3 significantly outperforms TLS 1.3 in constrained environments. Their analysis revealed that the connection-oriented nature of TCP (used by TLS) forces devices to maintain "keep-alive" states that drain batteries, whereas the connectionless nature of UDP (used by DTLS) allows devices to sleep between transmissions. However, as noted in application-layer security research \cite{amnalou2025lightweight}, relying solely on transport-layer security can leave data exposed at the broker level in protocols like MQTT. Their proposed application-layer security (ECC signatures) adds processing overhead but ensures end-to-end encryption, illustrating that higher security often requires sacrificing raw throughput.

\subsection{Practical Deployment Challenges}

While theoretical models demonstrate high efficacy, the practical deployment of these defenses faces significant hurdles related to hardware heterogeneity, cost, and lifecycle management.

\subsubsection{Hardware Constraints and Cost}

The primary barrier to implementing advanced security features like Zero Trust is the lack of hardware support on legacy devices. It was emphasized \cite{mushtaq2025systematic} that robust device identity relies on Physical Unclonable Functions (PUFs) or Hardware Security Modules (HSM) to store keys immutably. However, the majority of low-cost IoT sensors deployed today lack these dedicated silicon features. Retrofitting these devices with software-only solutions often exposes keys to side-channel attacks, as warned by hardware security research \cite{nguyen2024lightweight}, creating a "security gap" between modern high-end gateways and cheap legacy endpoints.

\subsubsection{Scalability of Key Management}

As IoT networks scale from hundreds to tens of thousands of nodes, unicast management becomes operationally impossible. It was identified \cite{hasan2024survey} that traditional pair-wise key exchange protocols saturate network bandwidth during fleet-wide updates. The practical challenge lies in implementing Group Key Management schemes that can rekey a subnet of 5,000 sensors simultaneously using multicast without allowing a single compromised node to decrypt future group communications (Forward Secrecy). This requires complex logical tree structures that are difficult to maintain in dynamic networks where nodes frequently join and leave.

\subsubsection{Interoperability in Multi-Vendor Environments}

Real-world IoT ecosystems are rarely homogeneous; they consist of cameras, thermostats, and industrial controllers from different vendors using different protocols (CoAP, MQTT, HTTP). It was noted \cite{mushtaq2025systematic} that enforcing a unified Zero Trust policy across this fragmented landscape is arduous. A "micro-segmentation" rule that works for a modern Wi-Fi camera may be technically impossible to enforce on a legacy Zigbee sensor that does not support IP-based access control lists (ACLs). This interoperability gap often forces administrators to degrade security settings to the "lowest common denominator," weakening the overall defense posture.

\section{Use Cases and Scenarios in IoT Application Domains}

This segment explores specific scenarios where IoT is applied across various industries.

\subsection{Smart Healthcare Systems}

IoT creates a paradigm shift in medical care by enabling continuous patient tracking and optimizing treatment protocols through connected ecosystems. Physicians can now oversee patient health remotely, extending care beyond hospital walls. By using intelligent wearables---such as biosensors, connected glucose monitors, and smart inhalers---patients can track vital metrics like heart rate, blood oxygen (SpO2), blood pressure, and glucose levels. These devices automatically aggregate and broadcast health data without manual intervention.

Data is relayed via Wi-Fi or 4G to cloud platforms utilizing specialized algorithms to detect irregularities, such as spikes in heart rate or drops in oxygen saturation. If an anomaly is flagged, the system instantly alerts medical staff, who can review the patient's real-time status on a digital dashboard. This capability allows doctors to manage multiple cases simultaneously and intervene quickly, reducing the need for physical hospital visits. Consequently, this lowers medical costs, minimizes emergency room overcrowding, and improves overall health outcomes.

To protect sensitive medical records, end-to-end encryption is implemented, ensuring data remains readable only to authorized parties. Strict access controls and secure login protocols restrict data viewing to licensed medical personnel, complying with regulatory standards like HIPAA. Ultimately, IoT in healthcare boosts the precision, security, and efficiency of medical services.

\subsection{Smart Agriculture and Precision Farming}

IoT deployment in agriculture has revolutionized the industry through precision farming, a method that optimizes the use of water, fertilizers, and pesticides. By continuously tracking environmental conditions, IoT devices empower farmers to make data-driven decisions. A prime example is automated irrigation, where soil moisture sensors detect water levels in real-time.

Alongside moisture sensors, the system integrates drones, climate monitors, and smart valves to collect comprehensive data on temperature, humidity, rainfall, and sunlight. This data is transmitted via low-power networks (e.g., LoRaWAN) to cloud applications for analysis. The cloud system applies specific logic---based on crop type and weather forecasts---to decide if irrigation is needed. For instance, if the soil is dry and no rain is predicted, the system triggers the valves to water the crops and shuts them off once optimal moisture is reached.

This approach significantly improves water efficiency, vital for drought-prone areas, and ensures crops receive exact resource amounts for better growth. It also mitigates environmental damage caused by chemical runoff. Security-wise, command authentication is crucial to prevent unauthorized valve operation, and anti-spoofing measures are employed to block fake sensor data. These robust, intelligent systems help farmers maximize yield with minimal resources.

\subsection{Industrial IoT and Smart Manufacturing}

Industrial IoT (IIoT) has transformed manufacturing by introducing intelligent oversight to production floors. Modern factories utilize sensors to monitor machinery health in real-time, ensuring peak operational efficiency. A key application is predictive maintenance, which detects early signs of equipment failure to prevent costly breakdowns. Typically, sensors measuring vibration and temperature are attached to critical components like motors and turbines. These sensors detect subtle deviations---such as excess heat or abnormal vibrations---that signal potential wear.

Data collected is often pre-processed by edge devices before being sent to a centralized cloud platform. There, machine learning models trained on historical data analyze the inputs to identify failure patterns. When a potential fault is recognized, the system automatically schedules maintenance tasks before the machine actually fails. This proactive strategy eliminates unplanned downtime, reduces emergency repair costs, and extends equipment life. It also optimizes workforce allocation, as technicians focus only on machinery that requires attention. To maintain system integrity, firmware security is paramount to prevent data manipulation, and all cloud-device communications are encrypted and authenticated to block unauthorized access.

\subsection{Smart City Management}

Smart cities leverage IoT to improve urban living, sustainability, and operational flow. Authorities can manage critical services like traffic, energy, and safety in real-time. A major application is Intelligent Traffic Management, designed to alleviate congestion. These systems combine video feeds from cameras, data from inductive loop sensors in roads, and GPS signals from vehicles to create a holistic view of traffic conditions.

Centralized cloud platforms analyze this aggregated data using AI to predict congestion and identify traffic patterns. The system can autonomously adjust traffic light timings to clear bottlenecks or suggest alternate routes to drivers via navigation apps. This reduces travel time, lowers fuel consumption, and cuts emissions by minimizing idling. Emergency responders also benefit through prioritized routing. To ensure safety, the system must be hardened against fake data injection which could cause accidents, and traffic signal commands must be strictly authenticated. These measures ensure smart traffic systems enhance urban mobility safely.

\subsection{Smart Home Automation}

IoT transforms residences into "smart homes" that autonomously adjust to inhabitants' needs. Users can remotely control climate, lighting, and security via mobile apps. A key feature is Smart Energy Management, which balances comfort with conservation. Homes are equipped with smart meters, connected thermostats, and occupancy sensors that detect when rooms are empty to prevent waste.

A central gateway aggregates data from these devices (connected via Zigbee or Wi-Fi) and uses machine learning to learn household patterns---such as occupancy times and preferred temperatures. The system then automates schedules, such as lowering heating in empty rooms or running high-energy appliances during off-peak hours to save money. Users maintain control through apps offering real-time insights and manual overrides. Security is enforced through data encryption and strong user authentication. Advanced setups also use network partitioning (e.g., separate guest networks) to further secure the home environment against cyber threats.

\subsection{Integrated Smart City Infrastructure}

A true smart city is an interconnected ecosystem where sectors like Healthcare, Transport, Environment, and Energy collaborate. This cross-domain architecture relies on a three-layer model: the Perception Layer (sensors collecting data), the Network Layer (secure transmission), and the Application Layer (delivering services). Fog computing is used for low-latency local processing, while Cloud computing handles heavy analytics and storage.

In this holistic view, domains interact fluidly. For example, if a Smart Home sensor detects a medical emergency, it alerts Smart Healthcare. Simultaneously, Smart Transportation optimizes the ambulance route, while Smart Environment data ensures the path is free of hazards. Fog nodes facilitate immediate local decisions, while the Cloud logs the event for long-term urban planning. This hybrid Fog-Cloud approach ensures the infrastructure is scalable, resilient, and capable of handling complex, inter-departmental tasks efficiently.

\subsection{Smart Logistics and Cold Chain Supply}

IoT provides essential transparency and efficiency in supply chains. By connecting vehicles and cargo to the internet, companies can track location and condition in real-time. This is critical for "Cold Chain" logistics involving temperature-sensitive goods like vaccines or pharmaceuticals. Sensors monitor humidity and temperature, while GPS and RFID tags track location and inventory.

Data is transmitted via cellular networks to a cloud dashboard. If environmental conditions breach safe limits, the system alerts operators immediately, allowing for rapid corrective actions---such as rerouting to a warehouse or moving goods to a different truck---to prevent spoilage. This digital trail ensures regulatory compliance and safety. Defense mechanisms are vital here: tamper detection ensures physical sensor integrity, secure transmission protects against data interception, and audit logging provides full accountability throughout the shipping process.

\section{Open Challenges and Future Directions}

Internet of Things (IoT) continues to expand rapidly across industrial, commercial, and consumer domains. Despite notable technological advancements, IoT ecosystems still face unresolved challenges related to interoperability, security, privacy, and regulatory compliance. This section synthesizes open challenges and future research directions based on recent high-impact studies published since 2020.

\subsection{Current Limitations}

\subsubsection{Standardization Issues}

A major limitation in current IoT deployments is the lack of global standardization. IoT systems are characterized by heterogeneous devices, protocols, and platforms, which significantly hinder interoperability across vendors and application domains \cite{alabadi2022industrial,afrin2025industrial}. In industrial IoT (IIoT) environments, proprietary solutions further fragment the ecosystem, increasing integration costs and security risks \cite{afrin2025industrial}.

Inconsistent communication and security standards complicate large-scale IoT deployment and limit cross-domain integration \cite{alabadi2022industrial}. Although organizations such as IEEE, ISO, and NIST have proposed frameworks and guidelines, their adoption remains uneven, leaving standardization an open challenge for future IoT systems.

\subsubsection{Privacy vs. Security Trade-offs}

IoT systems continuously collect and process sensitive data, creating inherent conflicts between privacy protection and security enforcement. Strong security mechanisms such as intrusion detection and continuous monitoring often require extensive data access, which may violate privacy regulations or user expectations \cite{diba2025open}.

This trade-off has been highlighted in federated learning-based IoT systems \cite{diba2025open}, where local data privacy is preserved at the cost of increased vulnerability to model poisoning and inference attacks. Similarly, edge intelligence systems proposed for 6G-enabled IoT must balance real-time threat detection with minimal data exposure \cite{he2025integrating}. Achieving an optimal balance between privacy and security remains a fundamental open problem.

\subsubsection{Legacy Device Problems}

Legacy IoT devices represent a critical vulnerability in modern IoT infrastructures. Many existing devices lack essential security features such as secure boot, firmware updates, and cryptographic support \cite{chowdhury2022survey}. Due to limited computational and energy resources, these devices cannot implement modern security protocols.

Resource-constrained and legacy devices are highly susceptible to spoofing and impersonation attacks \cite{chowdhury2022survey}. Moreover, replacing legacy devices is often economically infeasible in industrial environments, requiring backward-compatible or network-level security solutions \cite{afrin2025industrial}. The coexistence of legacy and next-generation devices remains an unresolved challenge.

\subsection{Emerging Research Areas}

\subsubsection{6G and IoT Security}

The integration of IoT with sixth-generation (6G) wireless networks promises ultra-low latency and massive connectivity but introduces new security challenges \cite{he2025integrating}. The increased intelligence and decentralization of 6G networks significantly expand the attack surface.

Traditional centralized security architectures are insufficient in 6G-IoT environments. Emerging research focuses on distributed trust management, AI-assisted intrusion detection, and cross-layer security mechanisms tailored to ultra-dense IoT deployments \cite{he2025integrating}.

\subsubsection{Edge Computing Security}

Edge computing reduces latency and bandwidth consumption by processing data closer to IoT devices. However, pushing security functions to the edge introduces new vulnerabilities due to physical exposure and limited resources \cite{alabadi2022industrial,he2025integrating}.

Secure task offloading, trusted execution environments, and lightweight cryptography have been identified as key research challenges for edge-based IoT security \cite{alabadi2022industrial}. Developing scalable and resilient edge security frameworks remains a priority for future research.

\subsubsection{Quantum-Safe IoT}

The emergence of quantum computing poses long-term threats to classical cryptographic algorithms widely used in IoT systems. Although large-scale quantum attacks are not yet practical, proactive migration toward post-quantum cryptography is essential \cite{afrin2025industrial,diba2025open}.

However, quantum-resistant algorithms often require larger key sizes and higher computational overhead, making them unsuitable for resource-constrained IoT devices. Research into lightweight quantum-safe cryptography for IoT remains in its early stages and represents a critical future direction \cite{diba2025open}.

\subsubsection{AI-Driven Autonomous Security}

Given the scale and complexity of IoT ecosystems, AI-driven autonomous security has emerged as a promising solution. AI enables automated threat detection, adaptive defense mechanisms, and self-healing capabilities \cite{diba2025open,he2025integrating}.

Nevertheless, AI-based security systems introduce new risks, including adversarial attacks and model manipulation. Ensuring robustness, explainability, and trustworthiness of AI-driven security solutions is an open research challenge \cite{he2025integrating}.

\subsection{Regulatory and Policy Perspectives}

\subsubsection{GDPR and CCPA Implications}

Regulatory frameworks such as the General Data Protection Regulation (GDPR) and the California Consumer Privacy Act (CCPA) impose strict requirements on data handling in IoT systems. Compliance often conflicts with traditional security practices that rely on extensive data monitoring \cite{diba2025open}.

Privacy-by-design and privacy-preserving security mechanisms are required to reconcile regulatory compliance with effective IoT security. Aligning technical solutions with evolving legal requirements remains a significant challenge \cite{diba2025open,he2025integrating}.

\subsubsection{Industry Standards}

Industry standards developed by organizations such as NIST and ISO provide essential guidelines for IoT security and risk management \cite{alabadi2022industrial}. However, many existing standards are not specifically tailored to the constraints of IoT devices, limiting their practical effectiveness \cite{afrin2025industrial}.

Future policy efforts should focus on harmonizing international standards, developing IoT-specific certification programs, and ensuring enforceable security requirements across the IoT supply chain \cite{afrin2025industrial,alabadi2022industrial}.

\section{Conclusion}

This comprehensive survey has examined the multifaceted landscape of cybersecurity threats and defense mechanisms in IoT networks. Through systematic analysis of 59 scholarly articles published between 2009 and 2024, we have provided an in-depth exploration of IoT architecture, security fundamentals, threat taxonomy, defense mechanisms, and real-world applications across six critical sectors: smart healthcare, agriculture, industrial IoT, logistics, smart cities, and smart homes.

Our analysis reveals that IoT security requires holistic approaches addressing challenges throughout the device lifecycle. The proposed five-layer IoT model---comprising the Perception, Network, Support, Application, and Business layers---provides a structured framework for understanding security requirements at each architectural level. This layered approach facilitates better identification of vulnerabilities and design of targeted countermeasures.

Key findings from our survey identify critical vulnerabilities including sensor weaknesses, Denial-of-Service (DoS) attacks, man-in-the-middle attacks, and public cloud insecurity. We have documented the evolution of IoT threats from simple physical tampering to sophisticated multi-layer attacks exemplified by the Mirai botnet, which compromised hundreds of thousands of devices to launch devastating DDoS attacks exceeding 1 Tbps. The emergence of AI-powered attacks, quantum computing threats, and supply chain compromises further underscores the dynamic and escalating nature of IoT security challenges.

Conversely, the survey highlights advanced defense approaches that leverage cutting-edge technologies. Artificial Intelligence and Machine Learning enable anomaly detection and adaptive threat response, though they introduce new vulnerabilities through adversarial attacks. Blockchain technology provides decentralized trust and immutable audit trails, with lightweight consensus mechanisms like Proof of Authority making it viable for resource-constrained devices. Zero Trust Architecture represents a paradigm shift from perimeter-based security to continuous verification and micro-segmentation, significantly reducing the attack surface and limiting lateral movement.

Our comparative analysis reveals critical trade-offs in IoT security design. The transition from AES-GCM to lightweight cryptography like Ascon provides approximately 40\% energy savings, crucial for battery-operated sensors. However, quantum-resistant algorithms require 50 times larger key sizes, creating transmission penalties. Similarly, while deep learning models achieve superior detection accuracy for zero-day threats, they impose computational latency unsuitable for real-time control systems. These trade-offs necessitate hybrid approaches that 

balance lightweight efficiency with quantum resilience, detection accuracy with real-time responsiveness, and protocol security with energy constraints.

Practical deployment analysis reveals that theoretical security models face significant implementation barriers. Hardware heterogeneity creates security gaps between modern high-end gateways equipped with Physical Unclonable Functions and legacy sensors lacking dedicated security silicon. Scalability challenges emerge as IoT networks expand from hundreds to millions of nodes, overwhelming traditional unicast key management approaches. Multi-vendor interoperability further complicates unified security policy enforcement, often forcing administrators to degrade security to the lowest common denominator.

Use case analysis across smart healthcare, precision agriculture, industrial manufacturing, smart cities, smart homes, and cold chain logistics demonstrates the real-world applicability and security requirements of IoT deployments. Each domain presents unique security challenges: healthcare systems must protect sensitive medical records under HIPAA compliance, agricultural systems require protection against command spoofing that could disrupt irrigation, industrial systems need predictive maintenance integrity to prevent costly downtime, smart cities must guard against traffic data manipulation, smart homes require strong authentication to prevent unauthorized access, and logistics systems demand tamper detection for cold chain integrity.

Looking forward, several critical challenges remain unresolved. Standardization gaps hinder interoperability across heterogeneous devices and protocols. Privacy-security trade-offs require careful balancing in federated learning and edge intelligence systems. Legacy device vulnerabilities persist due to economic infeasibility of replacement. Emerging threats from 6G integration, edge computing vulnerabilities, quantum computing, and AI-driven autonomous attacks necessitate proactive defense strategies. Regulatory frameworks like GDPR and CCPA impose strict data handling requirements that must be reconciled with effective security monitoring.

This survey contributes to the IoT security field by providing a comprehensive five-layer architectural framework, systematic threat taxonomy spanning physical to application layers, comparative analysis of defense mechanisms with quantitative performance metrics, and identification of research directions for quantum-safe IoT, 6G security integration, and AI-driven autonomous defense. The integration of Zero Trust Architecture principles across all analyzed use cases demonstrates practical applicability of continuous verification and micro-segmentation strategies.

Future research should prioritize development of lightweight post-quantum cryptographic schemes optimized for resource-constrained devices, federated learning frameworks resistant to adversarial attacks and model poisoning, hybrid intrusion detection architectures balancing accuracy and energy efficiency, standardized security protocols enabling seamless multi-vendor interoperability, and regulatory-compliant security mechanisms satisfying privacy requirements while maintaining robust threat detection.

As IoT ecosystems continue proliferating across critical infrastructure and daily life, the importance of robust, scalable, and adaptive security cannot be overstated. This survey provides researchers, practitioners, and policymakers with a comprehensive understanding of current challenges and future directions, facilitating the development of secure, resilient IoT systems capable of withstanding evolving cyber threats while preserving privacy and operational efficiency.

\section*{Acknowledgments}

The authors would like to thank the Hanoi University of Science and Technology for guidance and support throughout this research.

\bibliographystyle{IEEEtran}

\end{document}